\documentclass[prb,a4paper,twocolumn,floatfix,superscriptaddress,showkeys,amsmath,amssymb,nobibnotes,altaffilletter]{revtex4-1}

\usepackage{graphicx}
\usepackage{pslatex}
\usepackage{xspace}
\usepackage{subfigure}

\begin{document}

\preprint{}

\title{Spectroscopic Signatures of Photogenerated Radical Anions in Polymer-[C$_{70}$]Fullerene Bulk Heterojunctions}

\author{M. Liedtke}%
\affiliation{Experimental Physics VI, Julius-Maximilians-University of W{\"u}rzburg, 97074 W{\"u}rzburg, Germany}
\affiliation{Bavarian Centre for Applied Energy Research (ZAE Bayern), 97074 W{\"u}rzburg, Germany}

\author{A. Sperlich}%
\author{H. Kraus}%
\author{C. Deibel}%
\affiliation{Experimental Physics VI, Julius-Maximilians-University of W{\"u}rzburg, 97074 W{\"u}rzburg, Germany}

\author{V. Dyakonov}\email{dyakonov@physik.uni-wuerzburg.de}%
\affiliation{Experimental Physics VI, Julius-Maximilians-University of W{\"u}rzburg, 97074 W{\"u}rzburg, Germany}%
\affiliation{Bavarian Centre for Applied Energy Research (ZAE Bayern), 97074 W{\"u}rzburg, Germany}

\author{S. Fillipone}%
\affiliation{Departamento de Qu\'{\i}mica Organica, Universidad Complutense de Madrid, 28040 Madrid, Spain}%

\author{J. L. Delgado}%
\affiliation{IMDEA-Nanociencia, Facultad de Ciencias, Ciudad Universitaria de Cantoblanco, }%

\author{N. Mart\'{\i}n}%
\affiliation{Departamento de Qu\'{\i}mica Organica, Universidad Complutense de Madrid, 28040 Madrid, Spain}%
\affiliation{IMDEA-Nanociencia, Facultad de Ciencias, Ciudad Universitaria de Cantoblanco, }%

\author{O. G. Poluektov}%
\affiliation{Chemical Sciences and Engineering Division, Argonne National Laboratory, Argonne, Illinois 60439, USA}%

\date{\today}

\begin{abstract}
Light induced polarons in solid films of polymer-fullerene blends were studied by applying photoluminescence (PL), photo induced absorption (PIA) techniques as well as electron spin resonance (ESR). The materials used were poly(3-hexylthiophene) (P3HT) and poly-[2-methoxy, 5-(2'-ethyl-hexyloxy) phenylene vinylene] (MEH-PPV) as donors. As acceptors we used [6,6]-phenyl-C$_{61}$-butyric acid methyl ester ([C$_{60}$]PCBM) and various soluble C$_{70}$-derivates: [C$_{70}$]PCBM, diphenylmethano[70]fullerene oligoether (C$_{70}$-DPM-OE), C$_{70}$-DPM-OE2, and two fullerene dimers, C$_{70}$-C$_{70}$ and C$_{60}$-C$_{70}$ (all shown in figure~\ref{fig:1}). In all blends containing C$_{70}$ we found typical signatures which were absent if [C$_{60}$]PCBM was used as acceptor. Light-induced ESR revealed signals at $g\ge$2.005, which we previously assigned to an electron localized on the C$_{70}$ cage, the PIA measurements showed a new sub-bandgap absorption band at 0.92 eV, which we correspondingly ascribe to C$_{70}$ radical anions formed in the course of photoinduced electron transfer from donor to acceptor.
\end{abstract}



\maketitle

\section{Introduction}

Due to their potentially low manufacturing costs, ultra-thin film organic solar cells (OSC) may become highly competitive in the area of direct solar energy conversion. Efficiencies of about 8~\% have been reported recently [1] and additional research efforts are going to push that level further.

An important task along that way is to gain a better understanding of the processes taking place in the device during the conversion of incident light into electrical energy. Using blends of conjugated polymers and fullerenes, an efficient light induced charge separation can be achieved, subsequently followed by migration of the charges in the corresponding material phase of the blend to the device electrodes.
[C$_{60}$]PCBM, a soluble fullerene derivative, is used in the vast majority of the OSC devices reported. It is being outperformed recently by [C$_{70}$]PCBM, which is gaining attention due to its higher absorption in the visible part of the solar spectrum (figure~\ref{fig:2}) but also due to slightly higher open circuit voltage in the device [2].

Although understanding of the elementary steps of efficient charge separation and stabilization of the separated charge carriers in the photovoltaic materials is a prerequisite for improving the efficiency of organic PV cells, very little is known on the photophysics in C$_{70}$ containing composites. So far VIS-NIR spectra of C$_{70}$ solution reduced by potassium in tetrahydrofuran as well as low-frequency (LF) ESR studies were previously mentioned in the literature [3]. Recently, detailed studies on fullerene C$_{70}$ based hetero- and homodimers report additional photo-induced absorption and LF-ESR features due to [C$_{70}$]fullerenes [4].

Here we relate on the photoinduced charge separation state in polymer:[C$_{70}$]fullerene blends. In addition to our earlier publication [5], in which we obtained information on the g-tensor and therefore on the symmetry of the electronic wave function via D- (140 GHz) and X-band (9 GHz) ESR measurements, we present results of the sub-gap photo-induced absorption spectroscopy determining the excitation energies of the involved excited state.


\section{Experimental}

\begin{figure}
	\includegraphics[width=8.0cm]{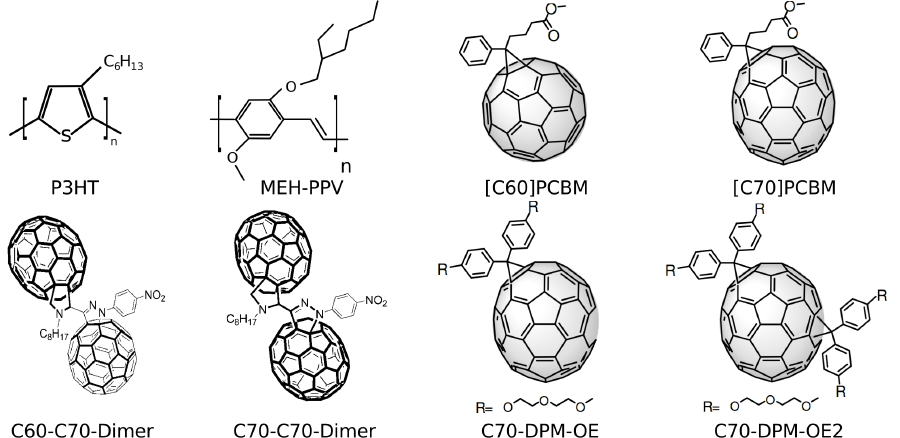}%
	\caption{Organic materials used in our measurements. The polymers P3HT and MEH-PPV worked as donors in our blend systems. All fullerene-derivates contain C$_{70}$ except for the [C$_{60}$]PCBM, which was used as reference material. %
	\label{fig:1}}
\end{figure}

P3HT (by Rieke Metals), PPV (by Sigma-Aldrich), [C$_{60}$]- and [C$_{70}$]PCBM (by Solenne B.V.) were used as purchased without any further purification. Details about the synthesis of the fullerene dimers are reported elsewhere [4].
For PIA spectroscopy the samples were mounted on a helium cold finger cryostat (15-293 K). During the measurement they were kept under dynamic vacuum to avoid photo-oxidation. The excitation source was a mechanically chopped 532 nm DPSS cw laser with a power of 45 mW. Additionally, cw illumination was provided by a halogen lamp. Both beams were superimposed on the sample. The transmitted light was collected by large diameter concave mirrors and focused into a cornerstone monochromator. Depending on the wavelength, the detection was provided by a silicon photodiode (550 nm-1100 nm), or by a liquid nitrogen cooled InSb-detector (1100 nm-5500 nm). Therefore, a broad energy range 0.23?2.25 eV (restricted by KBr cryostat windows) was accessible. The signals were recorded with a standard phase sensitive technique synchronised with the chopping frequency of the laser by using a Signal Recovery 7265 DSP Lock-In amplifier. Photoinduced variation of the transmission, $-\Delta T/T$ was monitored as a function of probe light wavelength. The samples were spin-coated from a 1:1 (weight ratio) polymer:fullerene solution onto a sapphire substrate under a nitrogen atmosphere.
Information about the ESR setup is given elsewhere [5].


\section{Results and Discussion}

\begin{figure}
	\includegraphics[width=8.0cm]{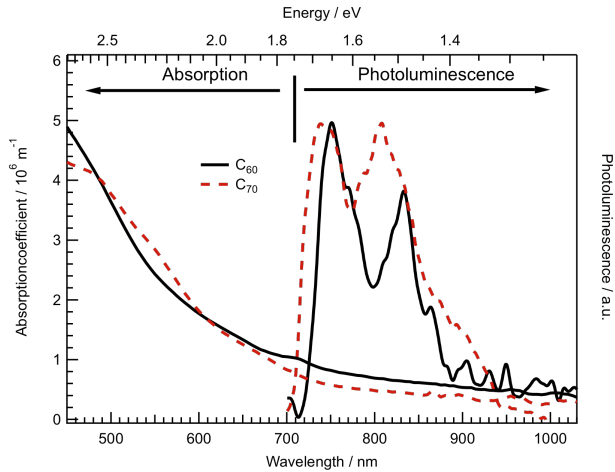}%
	\caption{Absorption and photoluminescence properties of [C$_{60}$]- and [C$_{70}$]-methanofullerene films. The absorption in the visible domain is enhanced for C$_{70}$ compared to C$_{60}$ by almost 20~\% at 550 nm. The vibronic structure of the photoluminescence of the fullerenes is visible, corresponding to the singlet transition energies. The absorption measurements were done at room temperature, while for the photoluminescence measurements the samples were cooled down to 30 K. %
	\label{fig:2}}
\end{figure}

Figure~\ref{fig:2} shows the absorption and photoluminescence spectra of C$_{60}$ and C$_{70}$ methanofullerenes in films. C$_{70}$ shows a increased absorption in the range from 500 nm to 600 nm (up to 20~\% at 550 nm) which is favorable as the sun spectrum in that range is quite strong and therefore organic solar cells using C$_{70}$ derivates as acceptors can utilize a higher fraction of incoming photons leading to a higher power conversion efficiency.

Films of C$_{60}$ and C$_{70}$ at low temperatures (30 K) emit a weak, yet detectable photoluminescence when excited by a green laser (532 nm). The spectra of both fullerene derivatives show a well resolved vibronic structure; the photoluminescence of C$_{70}$ being shifted to the blue by about 15 nm compared to C$_{60}$. The first two peaks of the C$_{70}$ vibronic structure are closer spaced, which indicates slightly higher Sx+1 $\leftarrow$ Sx transition energies for the C$_{70}$ molecule [8].

We studied the photophysical properties of different polymer:fullerene blends by applying PIA. The primary excitation by a green laser generates singlet excitons, which are separated at the polymer-fullerene interface yielding positive and negative charge carriers in the polymer and the fullerene phase respectively. After generation, they can be detected due to their typical optical absorption signatures using a white light continuum probe beam.

In many fullerene:polymer blends, transitions ascribed to polaronic states can be detected at 0.3eV and 1.2eV, while a pure polymer, such as P3HT shows a solitary peak at 1.05 eV of excitonic nature as no charge separation takes place. For P3HT or PPV blended with [C$_{60}$]PCBM these curves are well known and understood (6,12). 

\begin{figure}
	\includegraphics[width=8.0cm]{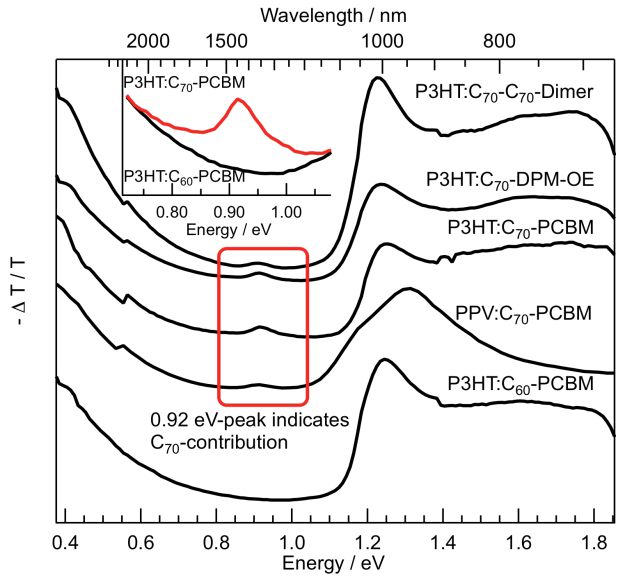}%
	\caption{PIA measurements for all used polymer:fullerene blends. While the structure of the spectra is similar, the peak at 0.92 eV only occurs in blends with [C$_{70}$]fullerenes but not with [C$_{60}$]PCBM as acceptor. The inset shows a detailed view of the spectra of [C$_{60}$]PCBM and [C$_{70}$]PCBM blended with P3HT. As the 0.92 eV peak occurs independently of the polymer used, it can be attributed to the C$_{70}$ anion. All measurements were made between  15 K - 30 K. %
	\label{fig:3}}
\end{figure}

\begin{figure}
	\includegraphics[width=8.0cm]{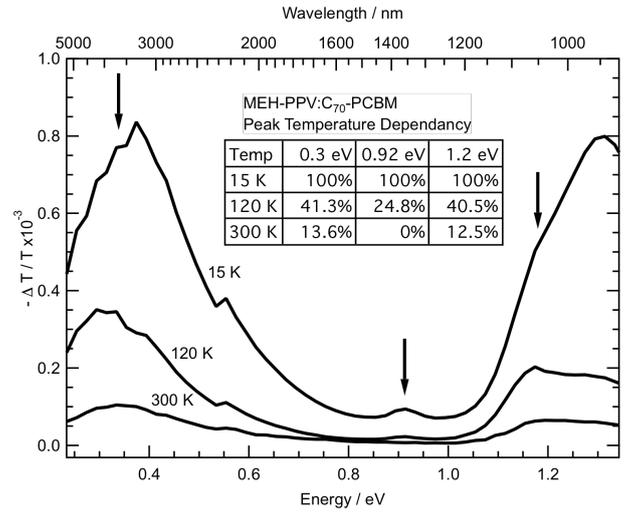}%
	\caption{PIA measurements of a MEH-PPV:[C$_{70}$]PCBM blend at different temperatures. The inset table shows the relative peak heights measured at the position of the three arrows in the spectrum. The temperature dependence is the same for the peaks at 0.3 eV and 1.2 eV (for 15 K we took the small shoulder as reference height), while the peak at 0.92 eV is decreasing stronger with temperature and vanishes at room temperature. %
	\label{fig:4}}
\end{figure}

In the PIA spectra of P3HT:[C$_{70}$]PCBM the same polaronic absorption bands can be detected at 0.3 eV and 1.2 eV, indicating efficient charge separation. Additionally we found a new absorption at 0.92 eV (figure~\ref{fig:3}), which has not been observed in C$_{60}$ based blends until now. This absorption peak is encountered in every [C$_{70}$] fullerene derivative blend investigated. As the examined fullerene derivatives here are equipped with different side chains, we can exclude these as a possible origin for the 0.92 eV peak. Furthermore, we exchanged the P3HT with other polymers, such as MEH-PPV, to verify whether or not the peak is related to the interaction between P3HT and C$_{70}$. As can be seen in figure~\ref{fig:4}, the peak at 0.92eV occurs in every blend containing a C$_{70}$-fullerene, independent of the polymer used or the side chains of the acceptor. As a reference, a P3HT:[C$_{60}$]PCBM PIA spectrum is shown as well. The peak at 0.92 eV is clearly absent in this blend. 

All spectra contain two pronounced transitions at 0.3 eV and 1.2 eV, due to the positive polaron created on the polymer, shifting slightly due to the varying energy levels of the different polymers. In the polymer-[C$_{60}$]PCBM blends, the signature of the C$_{60}$ anion radical peak is an interesting issue. It is missing in the PPV or P3HT based [C$_{60}$]PCBM blends. As discussed in literature [7], this may be due to the fact that the C$_{60}$ anion peak is at 1.2 eV and therefore superimposed by the stronger polymer polaron signal located at the same energy. We do observe an additional C$_{60}$ radical anion PIA peak in blends of C$_{60}$ and novel low-band polymers PTDBT confirming the finding in Ref [7]. More details about the fullerene anion signatures and the reasons behind the different energies of these will be published elsewhere. 

During photoinduced electron transfer between polymer and fullerene, a positive polaron and a negative radical anion is created. The positive polymer polaron is well known to show the characteristic 1.2 eV and 0.3 eV signatures, but does not explain the feature appearing at 0.92 eV. According to our findings, we associate this new feature to the negative radical anion on the [C$_{70}$]fullerene.
The conclusion of a different nature of the 0.92 eV is also consistent with the temperature dependence of the PIA peaks in the blend of MEH-PPV:[C$_{70}$]PCBM (figure~\ref{fig:4}). While the cation peaks at 1.2 eV and 0.3eV decrease from 100~\% (15 K) to 41~\% (120 K) and still show 13~\% peak height at room temperature, the peak at 0.92 eV is not measureable at room temperature; at 120 K it shows only 25~\% of the 15 K signal height. For this the broad background has been subtracted at the 0.92 eV peak position. 
We also discuss the possibility of a charge transfer (CT) state as origin of the additional absorption at 0.92 eV. The CT state occurs in blends during charge carrier generation, so it could also lead to the observed feature. But as the two polymers used here, MEH-PPV and P3HT, have different highest occupied molecular orbital (HOMO) energies (9, 10), it is highly unlikely for the CT state to have exactly the same energy in both blends [11].

\begin{figure}
	\includegraphics[width=8.0cm]{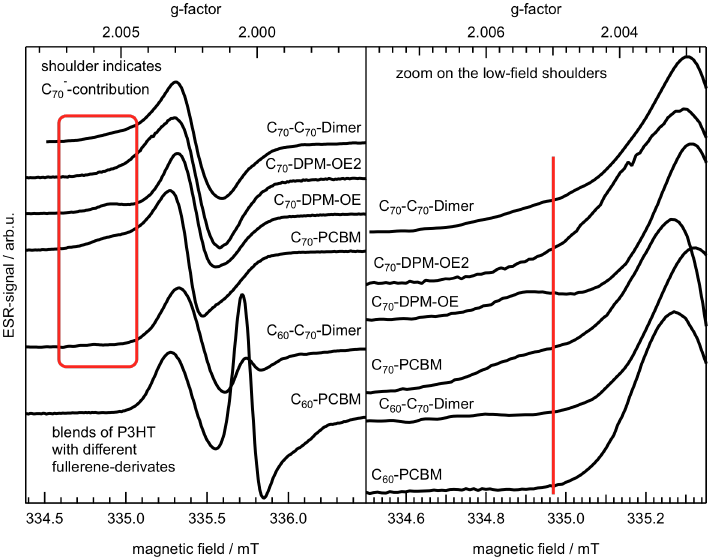}%
	\caption{X-Band ESR spectra of different P3HT:[C$_{70}$]Fullerene blends at temperatures of 30 K - 100 K, illuminated by a 532 nm laser. The main 1st derivative signal at $g=$2.002 is originating from positive polarons on the polymer. P3HT in blend with [C$_{60}$]PCBM is added for comparison, showing the same polaron signal at $g=$2.002 and the known spectrum of C$_{60}$ anions at $g=$2.000. All C$_{70}$-composites show an additional small low field shoulder for $g\ge$2.005 (indicated by the red rectangle in the left picture) that is assigned to the C$_{70}$ anion. The right picture shows a zoom on the low field shoulder. The red vertical line marks the lower limit (mT) of the polymer polaron spectrum.  %
	\label{fig:5}}
\end{figure}

In our recent work we identified spin signatures of C$_{70}$ related radical anions in blends with P3HT and reconstructed its g-tensor components [5]. Here we extend this study by investigating additional C$_{70}$ fullerene derivatives and dimers. In figure~\ref{fig:5} the light induced ESR spectra of P3HT blended with these fullerenes are shown. The spectra are shown in 1st derivative due to the detection method of the setup. Illumination was provided by a 532 nm DPSS laser and the samples were cooled to 30 - 100 K. In this temperature range only the ESR line intensities vary, but the line shapes remain almost identical, therefore the spectra can be shown together for comparison after normalization at the peak at $g=$2.003. All spectra show the typical 1st derivative absorption line at $g=$2.002, originating from positive polarons on the polymer. In P3HT:[C$_{60}$]PCBM blends, the negative anion radical of C$_{60}$ can be detected at $g=$2.000. For blends only consisting of P3HT and C$_{70}$-derivatives no such second signal can be identified immediately, because the C$_{70}$ radical anion line is strongly superimposed with the polymer polaron signal [5]. In the right half of figure~\ref{fig:5},  zooming on the low field fraction of the ESR reveals that all C$_{70}$-containing blends disclose a shoulder at $g\ge$2.005, while the spectrum of P3HT:[C$_{60}$]PCBM has no discernible signal at this position. Knowing all involved g-tensor components [5], we can relate this shoulder to the C$_{70}$ radical anion. Interestingly, the P3HT:[C$_{60}$-C$_{70}$] Dimer does not only show the weak $g\ge$2.005 shoulder, but also has a feature at $g=$2.000 resulting from the C$_{60}$ anion, combining both anion signals in one spectrum. This opens up a new interesting outlook of using fullerene heterodimers as model systems for studying spin?spin interactions

\section{Conclusions}

To summarize, we have studied a variety of C$_{70}$-fullerene derivates and dimers blended with different polymers, with the emphasis being on spectroscopic signatures of the photogenerated radical anion. We identified common C$_{70}$-related features in PIA and ESR spectroscopy, namely an additional sub-bandgap PIA peak at 0.92 eV and an ESR-shoulder at $g\ge$2.005. By comparing results from several different blends we can infer that these features are indeed signatures of the C$_{70}$ radical anion and not originating from the side chains of the fullerenes or CT-states between C$_{70}$-fullerenes and polymers. As anticipated, ESR signatures related to the C$_{70}$ radical anion were found and the corresponding g-values were in good agreement with our previous findings [5]. The obtained signatures are of importance, as they offer a measurable indicator of efficient charge transfer from the donor to the C$_{70}$ acceptor in blends used in organic solar cells.

\section*{Acknowledgments}

The work at the University of W{\"u}rzburg was supported by the German Research Foundation, DFG, within the SPP 1355 `Elementary processes in organic photovoltaics', under contract DY18/6-1. The MICINN of Spain (project CT2008-00795/BQU, R\&C program, and Consolider-Ingenio 2010C-07-25200) and the CAM (project P-PPQ-000225-0505) are also acknowledged. J. L. D. thanks MICINN of Spain for a Ramon y Cajal contract cofinanced by the EU social funds. OGP acknowledge support of the ANSER, an Energy Frontier Research Center funded by the U.S. Department of Energy, Office of Science, Office of Basic Energy Sciences.

\end{document}